\documentclass[sigconf]{acmart}
\usepackage{todonotes}
\usepackage{amsmath}
\usepackage{amssymb}
\usepackage{subcaption}
\usepackage{footmisc}
\usepackage{url}
\usepackage{fancybox}
\usepackage{multirow}
\usepackage{caption}
\usepackage{subcaption}
\def\BibTeX{{\rm B\kern-.05em{\sc i\kern-.025em b}\kern-.08emT\kern-.1667em\lower.7ex\hbox{E}\kern-.125emX}}
\newcommand{\specialcellL}[2][l]{%
  \begin{tabular}[#1]{@{}l@{}}#2\end{tabular}}
\newcommand{\specialcellC}[2][c]{%
  \begin{tabular}[#1]{@{}c@{}}#2\end{tabular}}

\setcopyright{none}

\copyrightyear{2019} 
\acmYear{2019} 
\acmConference[CIKM '19]{The 28th ACM International Conference on Information and Knowledge Management}{November 3--7, 2019}{Beijing, China}
\acmBooktitle{The 28th ACM International Conference on Information and Knowledge Management (CIKM '19), November 3--7, 2019, Beijing, China}
\acmPrice{15.00}
\acmDOI{10.1145/3357384.3358037}
\acmISBN{978-1-4503-6976-3/19/11}

\begin{document}

\title[Quantifying and Modeling Intra-community Conflicts in Online Discussion]{Into the Battlefield: Quantifying and Modeling Intra-community Conflicts in Online Discussion}

%

\author{Subhabrata Dutta, Dipankar Das}
\affiliation{\institution{Jadavpur University}
  \city{Kolkata, India}
  }
\email{{subha0009,dipankar.dipnil2005}@gmail.com}

\author{Gunkirat Kaur, Shreyans Mongia, Arpan Mukherjee, Tanmoy Chakraborty}
\affiliation{
  \institution{IIIT-Delhi, India}
  }
\email{{gunkirat15032,shreyans15178,arpan17007,tanmoy}@iiitd.ac.in}

%
\renewcommand{\shortauthors}{Dutta et al.}

%
\begin{abstract}
Over the last decade, online forums have become  primary news sources for readers around the globe, and social media platforms are the space where these news forums find most of their audience and engagement. Our particular focus in this paper is to study conflict dynamics over online news articles in Reddit, one of the most popular online discussion platforms. We choose to study how conflicts develop around news inside a discussion community, the {\em r/news} subreddit. Mining the characteristics of these engagements often provide useful insights into the behavioral dynamics of large-scale human interactions. Such insights are useful for many reasons --  for news houses to improvise their publishing strategies and potential audience, for data analytics to get a better introspection over media engagement as well as for social media platforms to avoid unnecessary and perilous conflicts.

In this work, we present a novel quantification of conflict in online discussion.
Unlike previous studies on conflict dynamics, which model conflict as a binary phenomenon, our measure is continuous-valued, which we validate with manually annotated ratings. We address a two-way prediction task. Firstly, we predict the probable degree of conflict a news article will face from its audience. We employ multiple machine learning frameworks for this task using various features extracted from news articles. 
Secondly, given a pair of users and their interaction history, we predict if their future engagement will result in a conflict. We fuse textual and network-based features together using a support vector machine which achieves an AUC of 0.89. Moreover, we implement a graph convolutional model which exploits engagement histories of users to predict whether a pair of users who never met each other before will have a conflicting interaction, with an AUC of 0.69. 

We perform our studies on a massive discussion dataset crawled from the Reddit news community, containing over $41k$ news articles and $5.5$ million comments. Apart from the prediction tasks, our studies offer interesting insights on the conflict dynamics -- how users form clusters based on conflicting engagements, how different is the temporal nature of conflict over different online news forums, how is contribution of different language based features to induce conflict, etc. In short, our study paves the way towards new methods of exploration and modeling of conflict dynamics inside online discussion communities.

\end{abstract}

\maketitle
\section{Introduction}
\label{sec:intro}
Mining knowledge from social media has gained tremendous attention among the research community in recent years. Endeavours started with entity recognition, opinion mining, object detection, etc.; current advents are pushing barriers to more complex analysis such as influence detection, malicious activity identification, multi-modal and heterogeneous data mining, etc. Usage of social media platforms are now so all-encompassing that these analyses yield rich insight into individual and community interaction in general. Online discussion forums are a particular type of social media and networks, which, due to their ever growing usage and popularity, need no introduction today. Reddit\footnote{\url{www.reddit.com/}} is one such example, where people across the globe engage in discussion related to  innumerable sets of topics.

With more and more people coming together in this virtual world, differences of opinions and conflict are an inevitability. Conflict may arise from several premises -- partial knowledge, socio-political understandings, clash of cultural and moral positions, and many more. It can be raised and developed from purely virtual individual interactions as well as real-world happenings. Although any difference of opinion can be identified as a conflict, its actual aspects are versatile. It may manifest itself within a vast spectrum, from constructive debates with well-formed argumentation to degenerated, unhealthy cyber-bullying and abuse. Thus a better introspection into the complex dynamics of conflict over online discussion platforms may provide more useful insights to the data analytics and social computing community  as well as help moderators of online platforms to identify and eliminate abusive conflicts and make the web a better place.

Versatility in manifestation of conflict is also the primary challenge of modeling conflict dynamics. Let us take the following three comments taken from Reddit:\\\\
\shadowbox{\begin{minipage}[t]{.95\columnwidth}

{\bf Comment 1:} I'm talking specifically about the 2010 Afghan War Diary, when Wikileaks was too lazy to scrub the names of about 100 Afghan civilian informants, thus revealing their identities to Taliban death squads. You actually sound a lot like Assange, who when asked why he didn't bother scrubbing the names said ``Well, they're informants, so, if they get killed, they've got it coming to them. They deserve it.''
\end{minipage}}\\
\shadowbox{\begin{minipage}[t]{.95\columnwidth}

{\bf Comment 2:} Assange didn't put them in danger. Participating in an illegal war and murdering innocent people in a country that never attacked the US put them in danger. Being actual Nazis put them in danger. Anyone who does that deserves to have a light shone on what they are doing, so that they hopefully stop. So sorry your conscience is worked up over maybes, instead of the hard reality of all the ACTUAL MURDERING that the US committed.
\end{minipage}}\\
\shadowbox{\begin{minipage}[t]{0.95\columnwidth}

{\bf Comment 3:} You seem a bit daft. If you were in Vichy France, informing for the Nazis, do you think you would have an expectation of privacy?When people like you start owning up to who the real monsters are, then the world can change.
\end{minipage}}

Both comments 2 and 3 are put in reply to comment 1, and both of them hold an opposite view. But how do we decide which one is more conflicting? Comment 3 is more subjectively aggressive towards the user posting comment 1. However, if we look at the content, comment 2 presents an opposite opinion in a more profound sense. Previous studies \cite{kumar2018community} on conflicts either treated it as a binary phenomenon; or identified controversy scores over topics and not between two text segments \cite{garimella2018quantifying}. Sophisticated NLP tools may come handy in this content; however one major downside is their lack of scalability in handling large-scale online data. In this work, we focus more on objective, argumentative conflict, rather than subjective, aggressive conflict. Simply put, we define comment 2 to hold more conflicting opinion compared to comment 1.

Online discussion platforms, through the lenses of engagement conflict, becomes a more complex dynamical process when the system interacts frequently with external sources. In this work, the external source is online news. Reddit has a specific community, \textit{r/news}, dedicated to discuss on news articles from various online news sources. Users post their views regarding news report and are engaged into discussion. Here a two way conflict comes into play -- users holding opposite opinions against a report and users holding opposite opinions towards each other. These two conflicts are even related; previous studies showed that certain news reports tend to blow up conflict of opinion between readers, mostly due to the topic of the news, language usage, political bias, etc. \cite{meyers1994defining,cramer2011controversy}.

\begin{figure}
    \centering
    \includegraphics[width=\columnwidth]{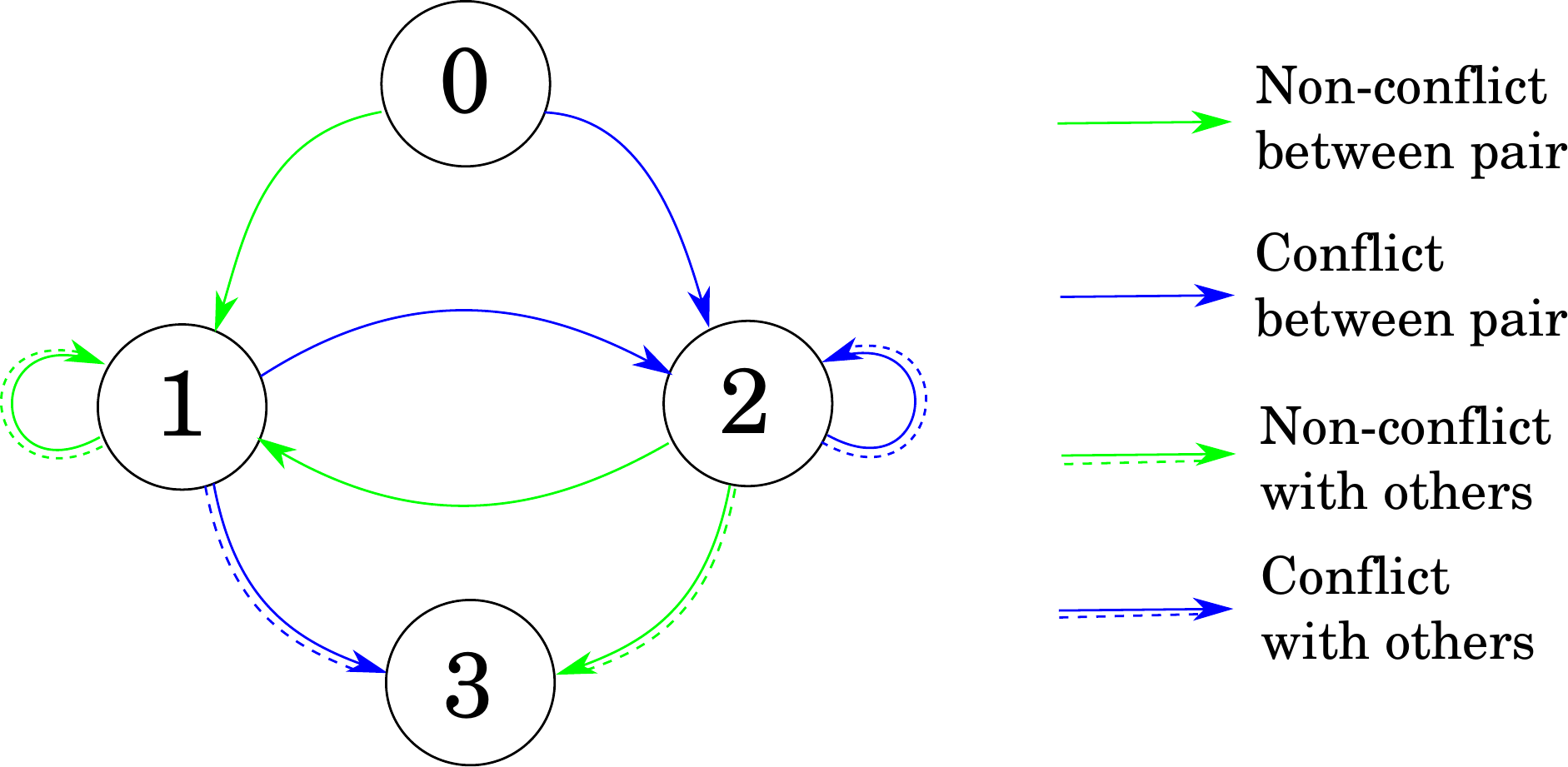}
    \caption{Hypothetical state-transition model of conflict for pair of users; state $0$ signifies starting of engagement between a hypothetical user pair.}
    \label{fig:conflict_state}
\end{figure}

\begin{table}
    \centering
    \small
    \begin{tabular}{|l|l|}
    \hline
         {\bf Notation} & {\bf Denotation}  \\
         \hline
         $T$ & Corpus-wide keyword set\\
         $T_D$ & Keywords present in document $D$\\
         $TS_D$ & Target-sentiment vector of document $D$\\
         $TS_u$ & \specialcellL{Target-sentiment vector averaged over\\ comments from user $u$}\\
         $N_i^k$ & \specialcellL{No. of comments from user $u_i$\\ containing term $T[k]$}\\
         $cf(D_1,D_2)$ & Conflict score between documents $D_1,D_2$\\
         $nc(N)$ & Total conflict towards news article $N$\\
         $G'(t)$ & Dynamic user engagement network\\
         
         \hline
         
    \end{tabular}
    \caption{Important notations used throughout the paper.}
    \label{tab:notation}
    \vspace{-5mm}
\end{table}

The state transition model in Figure \ref{fig:conflict_state} can be hypothesized as an abstract model of inter-user conflict dynamics. A transition from 0 to 1 or 2 signifies interaction between a hypothetical user pair. Any state transition from 1 or 2 can be of two types: either the users interact with each other (solid lines) or they interact with rest of the users (solid + dashed lines). Then state 1 corresponds to users having only non-conflicting engagement with each other, while state 2 denotes only conflicting engagements. State 3, in either way, identifies user groups who have preferential conflict. 

This abstract model can be actuated with a dynamic user-user interaction graph, with edges between users signifying previous interactions. We can further weigh these edges according to the degree of conflict arose in previous interactions. The problem of predicting future conflict between any two users then translates into a signed link prediction task. 

Our contributions in this work are as follows:
\begin{enumerate}
    \item We define a simple yet powerful and scalable measurement of conflict between pair of documents, focused on objective expression of opinion. We use target dependent sentiment scoring to compute a continuous valued score between text documents. We use this metric to quantify conflict between news reports and their audience as well as between user pairs interacting over discussion comments, on a large  dataset of news articles and corresponding discussions from Reddit \textit{r/news}. We manually annotate randomly selected news report-comment pair and comment-comment pair to test our conflict metric. We achieve $0.96$ and $0.79$ mean squared error over $[0-10]$ interval of conflict rating. For ranking comments according to the conflict they express towards particular news reports and comments, our method achieves mean average precision of $0.77$ and $0.83$, respectively.
    \item Using the conflict measurement, we attempt to predict the degree of conflict a news article will experience from the users reading and discussing it. Our prediction is solely based on the content of the article. We extract several textual features from the articles and employ multiple machine learning algorithms. We achieve symmetric mean average percentage error of $0.077$ with  with Support Vector Regression model.
    \item We attempt to predict whether a future interaction between any two given users will be conflicting or not, given their previous history of comments and engagement. We implement a Support Vector Machine based framework with selected textual and network-based features for this task, which achieved 0.89 AUC. We perform a fusion of textual features extracted from users' comment history and their interaction feature over the engagement network using graph convolution over dynamic user-user engagement, which correctly predicts conflict type between users (who have no previous history of interaction) with 0.69 AUC. 
    \item We conduct several experiments using the conflict metric to reveal intriguing patterns of conflict dynamics of news reporting over \textit{r/news} community. We explore how conflict towards news articles from different online news sources vary over time, and different news sources trigger inter-user conflict at different degrees. 
    \item We explore how inter-user conflict patterns emerge over time in discussion threads as well as in interaction network. We identify different community formation through conflict, which closely follow the abstract conflict model we described in Figure \ref{fig:conflict_state}.
\end{enumerate}


\section{Related works}
\label{sec:related_work}

In this section, we describe previous studies which we deem to be closely related to our work.

{\bf Conflict in community interaction}, which is the prime theme of our study, is a well studied problem in social network theory, psychology and sociology \cite{zachary1977information,nelson1989strength}. Different models and valuable introspection have emerged from these studies, such as how people tend to adapt towards certain acquaintances after initial conflict, fission in small group networks post conflict, emotional effects of conflict on individuals etc. However, studies on its online counterparts are much recent. Most of the studies in controversy  and polarization over social media are based on Twitter \cite{conover2011political,garimella2017long}. Garimella et al. \cite{garimella2018quantifying} proposed a graph-based approach to identify controversial topics on Twitter and measures to quantify controversy of a topic. They used 20 different hashtags to classify topics of conversation. Partitioning retweet, follow and reply graphs they compute the controversy related to each topic. Their work suggested the inefficiency of content-based measurements of controversy, majorly attributed by short spans of texts in tweets and high noise. Guerra et al. \cite{guerra2013measure} proposed a similar approach to measure polarization over social media; there data also is mostly based on Twitter. However, one must keep in mind that, the nature of conflict for microblogs is substantially different from that of discussion forums, primarily due to the size of the text. Kumar et al. \cite{kumar2018community} focused on Reddit to identify roles of conflict in community interactions. They performed their study on 36,000 Reddit communities (subreddits), identifying relation between inter-community mobilization and conflict. Their study also includes patterns of how people `gang up' on the verge of conflicting engagements. They predicted mobilizations between communities based on conflicts using user-level, community-level and text-level features. 
They achieved 0.67, 0.72 and 0.76 AUC using Random Forest, LSTM and ensemble of both, respectively. 
Our work can be thought of as another side of their story -- while they focused on conflict as a inter-community phenomenon, we attempt to address its dynamics in a microscopic level, inside a single community.

{\bf Stance detection and opinion mining} is closely related to conflict identification and measurement. Most of the previous works in stance detection are based on stance classification of rumors in Twitter \cite{zubiaga2018discourse,lukasik2016hawkes,mohammad2016semeval}. Rosenthal and McKewon \cite{rosenthal2015couldn} propsed a agreement-disagreement identification framework for discussions in Create Debate and Wikipedia Talkpages. They defined various lexical and semantic features from discussion comments and achieved an average accuracy of 77\% on the Create Debate corpus. Zhang et al. \cite{zhang2017characterizing} used discourse act classification on Reddit discussions to characterize agreement-disagreement over discussion threads. Dutta et al. \cite{dutta2019did} employed an attention-based hierarchical LSTM model for further improvement of discourse act classification on the same dataset.

{\bf News popularity prediction}, though does not handle conflict explicitly, is related to this work as it deals with engagement dynamics of online news. Previous studies can be classified into two main heads of approach -- popularity of news in social media platforms \cite{wu2015analyzing,rizos2016predicting,piotrkowicz2017headlines}, and popularity of news on web in general \cite{fernandes2015proactive,keneshloo2016predicting}. The second approach deals with the prediction problem unaware of inter-user network information, thereby excludes the explicit interaction of users with themselves and with the news sources. Popularity prediction models focus only on the degree of engagement a news gets, without concerning about the types of engagement, which is our focus in this work.

{\bf Link prediction on social networks}, as we already stated, is closely related to our formulated problem of predicting future conflict between users. There is rich literature focusing on this task \cite{liben2007link,gilbert2009predicting,aiello2012friendship,wang2015link}. Bliss et al. \cite{bliss2014evolutionary} used evolutionary algorithm for link prediction in dynamic networks. One important advancement in recent times for learning graph-based data is Graph Convolution Networks \cite{kipf2016semi,defferrard2016convolutional}. Zhang and Chen \cite{zhang2018link} applied convolution on enclosing subgraphs for link prediction. Berg et al. \cite{berg2017graph} also defined recommendation as a link prediction problem and used graph auto-encoder using deep stacking of graph convolutional layers. 
\section{Data}
\label{sec:data}
We crawled discussion threads containing at least one news link in the  posts or comments from \textit{r/news} subreddit, starting from 2016-09-01 to 2019-01-16. Out of 43,343 discussion threads crawled, we discarded threads containing less than 10 comments. The remaining 17,351 threads containing a total of 5,502,258 comments were used for the experiments. We also crawled news articles mentioned in the threads, resulting in a total of 41,430 news articles from 5,175 different news sources.\footnote{We have made the dataset containing the news articles public.}

To evaluate our conflict measurement strategy, we employed three expert annotators\footnote{They were experts on social media and their age ranged between 25-40 years.} to identify conflict between two given texts (articles/comments). We asked them to rate an interaction with higher conflict score than another if they found more elaborate opposition in the first one.  We provided the annotators with multiple examples annotated by us (one of these examples is presented in Section \ref{sec:intro}). We asked them to annotate the conflict in $[0-10]$ scale such that non-conflicting and highly conflicting texts will receive $0$ and $10$, respectively. For any interaction where only negativity has been expressed (sarcasm, popular slang without mentioning to what or whom it is addressed), we asked the annotators to rate as 1. We compute final ratings as the average of the ratings received. A total of randomly selected $3,734$ news-comment pairs and $6,725$ comment-comment pairs were annotated. The inter-annotator agreement based on Fleiss' $\kappa$ \cite{fleiss1971measuring} is $0.79$.   

\section{conflict Quantification}
\label{conflict_quantify}
Given two text segments, we measure conflict between them as how much opposite sentiment they exhibit. Here, we use target-dependent sentiment  measurement (TD-sentiment) as sentence-level sentiment may not be a good indicator of stance towards a motion. Let us take the following two sentences:
\begin{enumerate}
    \item \textit{Applauds for the writer to rightly explain why immigration is not a real problem.}
    \item \textit{This is an extremely good analysis of why immigration should be stopped.}
\end{enumerate}
Both of these sentences have positive sentence-level sentiment, though they carry conflicting opinion towards immigration. TD-sentiment for the term `\textit{immigration}' is neutral for sentence 1 and negative for sentence 2. From this, we can conclude that these two sentences are potential indicator of conflict.

As defined in our problem statement, we compute conflict between news article and platform users as well as between pair of users. Firstly, we compute a set of keywords from our dataset (comments + news articles). We tag the sentences using Spacy\footnote{\url{https://spacy.io/usage/linguistic-features}} parts-of-speech tagger and collect nouns only, after removing stopwords and lemmatization. To handle co-references of persons, we substitute nominal pronouns `\textit{he}' and `\textit{she}' by the last named-entity found with `Person'-tag. We include all the named entities in our keyword set, and top 60\% of the rest, ranked in order of tf-idf values. This results in a final corpus-wide term set $T$.

Next, we compute TD-sentiment of news articles and comments using Multi-Task Target Dependent Sentiment
Classifier (MTTDSC), a  state-of-the-art deep learning framework proposed by
Gupta et al. \cite{gupta2019multi} recently. MTTDSC is informed by feature representation learnedd for the related auxiliary task of passage-level sentiment classification. For the auxiliary task and main task, it uses separated gated recurrent unit (GRU), and sends the respective states to the fully connected layer, trained for the respective task. The model is trained and evaluated using multiple manually annotated datasets \cite{twitter_corpus, wang2017tdparse,dong2014adaptive}. 

Let a document $D$ (a single comment or a news article) be a sequence of sentences $[s_1, s_2, \cdots, s_n]$ and $T_D\subset T$ be the keyword set present in $D$ (where $T$ is the corpus-wide term set defined earlier). For any $t\in T$   occurring in $s_i$, MTTDSC computes a three class probability (positive, negative and neutral) vector $v_t^i$. Then for all the occurrences of $t$ in $D$, we compute aggregate sentiment of $D$ towards $t$ as $S_{D,t} = argmax(\frac{1}{n}\sum_{i} v_t^i)$, $S_{D,t}\in \{1,2,3\}$, where negative, neutral and positive sentiments are represented by 1, 2 and 3 respectively. Following this, we construct a vector $TS_D$ of size $|T|$ such that,
\begin{equation}
\label{eq:td-senti-vector}
    TS_D[i] = 
    \begin{cases}
    S_{D,T[i]}& \text{if } T[i]\in T_D\\
    0 & \text{otherwise}
    \end{cases}
\end{equation}
 $TS_D$ now represents the aggregate sentiments of document $D$ towards all the terms present in it. For any two documents $D_1$ and $D_2$, we then compute the \textit{conflict factor} ($cf$) between them using their aggregate TD-sentiment vectors $TS_{D_1}$ and $TS_{D_2}$ as follows:
\begin{equation}
\label{eq:conflict_factor}
    cf(D_1, D_2) = \sum_{i=0}^{|T|}min(TS_{D_1}[i],TS_{D_2}[i],1)\lvert TS_{D_1}[i]-TS_{D_2}[i]\rvert
\end{equation}
The component $min(TS_{D_1}[i],TS_{D_2}[i],1)$ returns 0 when either of the $i^{th}$ terms of 
$TS_{D_1}$ and $TS_{D_2}$ are 0, i.e., the term is not common, and 1 otherwise. This excludes terms which are not present in either of the texts to contribute to conflict computation. The value of the component $\lvert TS_{D_1}[i]-TS_{D_2}[i]\rvert$ can be 0 (when both texts have same sentiment towards the term), 1 (when one of texts hold neutral sentiment and other one positive or negative) and 2 (when texts hold opposite sentiments).

\section{News-User Conflict Prediction}
\label{sec:news_conflict}

Given a news article $N$ and the set of all comments $C$ related to $N$, we define \textit{News Conflict Score} as,
\begin{equation}
\label{eq:news_conflict_score}
    nc(N) = \frac{1}{\lvert C\rvert}\sum_{c\in C}cf(N,c)
\end{equation}
This is a normalized score referring to what degree users oppose the views presented in the news article. We then extract following features from news texts to predict this score given a news article:
\begin{enumerate}
    \item {\bf TD-sentiment vector}, entity-wise sentiment expressed in the news, as we compute $TS_D$ in Eq.~\ref{eq:td-senti-vector}.
    
    \item {\bf Count of positive, negative and neutral words}, tagged using SenticNet  \cite{cambria2018senticnet}.
    
    \item {\bf Cumulative entropy of terms}, given by,
    \begin{equation*}
        p=\frac{1}{|T|}\sum_{t\in T}tf_t(\log|T|-\log(tf_t))
    \end{equation*}
    where $T$ is the set of all unique tokens in the corpus, and $tf_t$ is the frequency of term $t$ in the news text.
    
    \item {\bf Fraction of controversy and bias words}, measured using the lexicon sets General Inquirer\footnote{\url{http://www.wjh.harvard.edu/~inquirer/homecat.htm}} and Biased Language\footnote{\url{http://www.cs.cornell.edu/~cristian/Biased_language.html}}; we use the fractions of these lexicons present in the article as controversy and bias features.
    
    \item {\bf Latent semantic features} using ConceptNet Numberbatch pretrained word vectors\footnote{https://github.com/commonsense/conceptnet-numberbatch} \cite{speer2016ensemble}; we compute TF-IDF weighted average of the vectors of the words present in an article to represent latent semantics of the article.
    
    \item {\bf LIX readability} \cite{bjornsson1983readability}, computed as: 
         $r = \frac{|w|}{|s|}+100\times\frac{|cw|}{|w|}$,
     where $w$ and $s$ are the sets of words and sentences respectively, and $cw$ is the set of words with more than six characters. Higher value of $r$ indicates harness of the users to read the article. 
     
     \item{\bf Gunning Fog} \cite{gunning1969fog}, computed by: $0.4\times(ASL + PCW)$, where  ASL is the average sentence length, and PCW is the percentage of complex words. Higher value of this index indicates harness of the users to read the article.
     \item{\bf Subjectivity}, calculated using TextBlob\footnote{\url{https://textblob.readthedocs.io/en/dev/}}. Its values lie in the range [0,1].
     
\end{enumerate}
To predict the conflict score $nc(N)$, we use three regression models: {\bf Lasso}, {\bf Random Forest Regressor}, and  {\bf Support Vector Regressor}. 

\section{Inter-user conflict prediction}
As already stated, we define the inter-user conflict prediction as a binary classification task to decide whether two users will engage in a conflict given their previous engagement history. We represent engagement history as a weighted undirected graph $G = \{V, E, W\}$, where every node $v_i\in V$ represents a user $u_i$, and every edge $e_{ij}\in E$ connects two nodes $v_i, v_j$ if and only if $u_i$ and $u_j$ have been engaged with each other earlier (i.e., either of them have commented in reply to at least one comment/post put by other). Every edge $e_{ij}$ is accompanied by a weight $w_{ij}\in W$ equal to the average conflict between $u_i$, and  $u_j$, which is computed as follows:
\begin{equation}
    w_{ij} = \frac{1}{N_{ij}}\sum_{k=0}^Ncf(D_{k}^{i},D_{k}^{j})
\end{equation}
where $D_{k}^{i}$ and $D_{k}^{j}$ represent the comments posted by $u_i$ and $u_j$, respectively at their $k^{th}$ interaction, and $N_{ij}$ is the total number of such interactions already occurred. $cf(D_{k}^{i},D_{k}^{j})$ is computed following Eq.~\ref{eq:conflict_factor}.  

To predict conflict between user pairs, we propose four different frameworks: one using graph convolution and three using Support Vector Machine (SVM) with different feature combinations. 

\subsection{Graph convolution on engagement network}
\label{subsec:gcn}
As typical user-user engagement networks of online discussion platforms are huge in size, we need to implement graph convolution over a subgraph. To predict the engagement type between a pair of users corresponding to vertices $v_i$ and $v_j$, we compute an enclosed subgraph $G_{sub} = \{V_{sub}, E_{sub}\}$ containing $v_i,v_j$ from $G$ such that $\forall v_k\in V_{sub}$, $dis(v_i,v_k),dis(v_j,v_k)\leq dis_{max}$, where $dis(v_i,v_k)$ is the length of the shortest path between $v_i$ and $v_k$, and $dis_{max}$ is a threshold distance (see Section \ref{sec:eval} for more details). All the edges in $E_{sub}$ share the same weight as in $G$.

We compute the adjacency matrix $\mathbf{A}$ from $G_{sub}$ as follows:
\begin{equation}
    \mathbf{A}[i][j] = \mathbf{A}[j][i] = 
    \begin{cases}
    w_{ij}& \text{if } e_{ij}\in E_{sub}\\
    0 & \text{otherwise}
    \end{cases}
\end{equation} 

We represent every node $v_i$ with a $d$-dimensional feature vector $x_i\in \mathbb{R}^d$, which represents previous commenting history of  user  $u_i$. We compute $x_i$ as the average over all the feature vectors corresponding to previous comments from $u_i$, using the same feature selection method as in Section \ref{sec:news_conflict} with an additional feature as follows -- a binary vector representing the news sources the user is engaged with. This leaves us with a tensor representation of user vertex features $\mathbf{X}=\{x_1,x_2,\dots,x_{\lvert V\rvert}\}$.

The adjacency matrix $\mathbf{A}$ and the vertex feature tensor $\mathbf{X}$ now represent network history and comment history of all the users at an instance, respectively. First, we learn a lower dimensional feature representation $\mathbf{X'}$ from $\mathbf{X}$ as follows:
\begin{equation}
    \mathbf{X'} = \sigma_{r}(\mathbf{K_f}^\top \mathbf{X} + \mathbf{B_f})
\end{equation}
where $\mathbf{K_f}$ and $\mathbf{B_f}$ are kernel and bias matrices to learn while training and $\sigma_{r}(x)=max(x,0)$.
\begin{figure}[!t]
    \centering
    \includegraphics[width=0.3\textwidth]{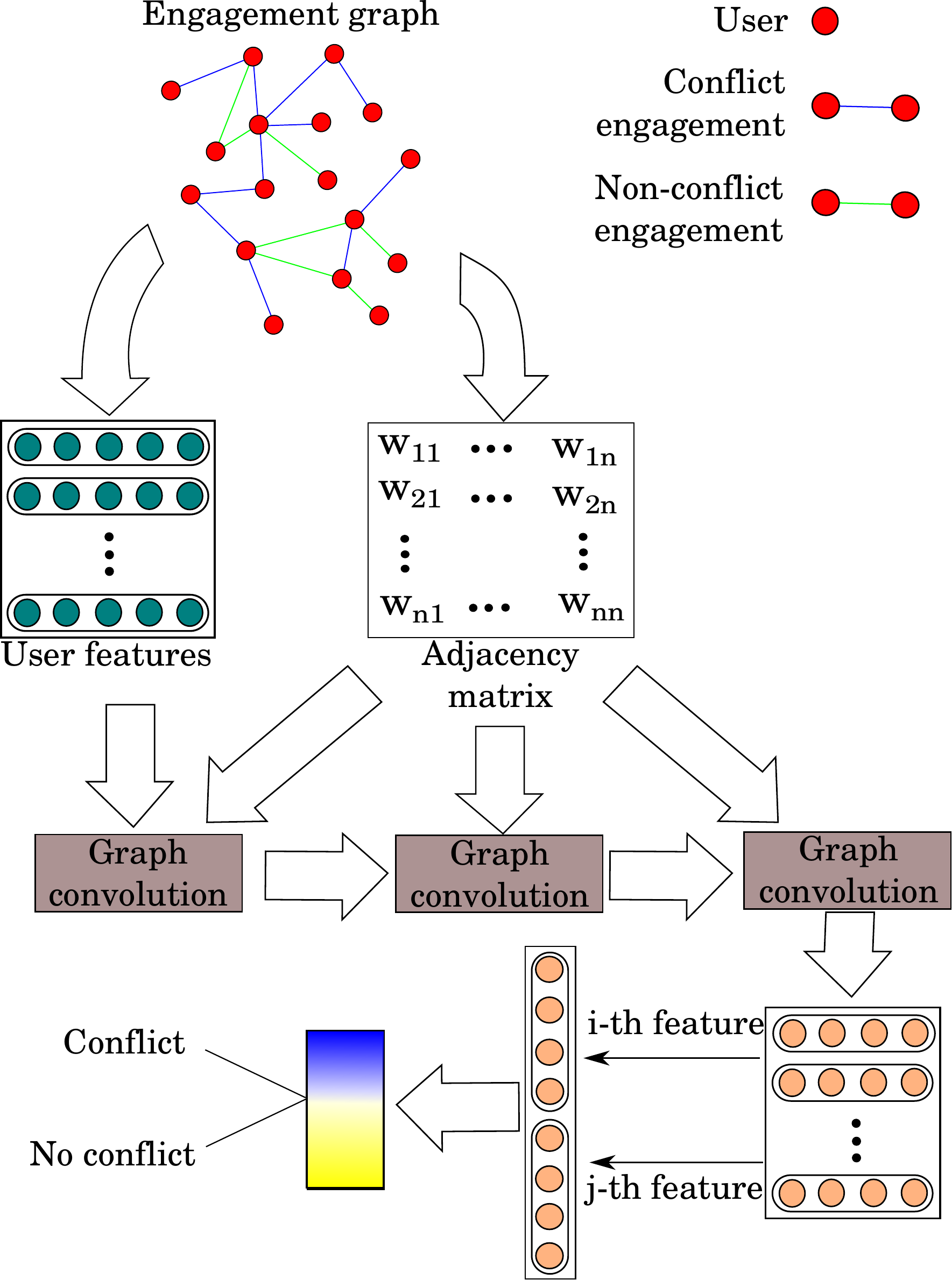}
    \caption{Inter-user conflict prediction using graph convolution.}
    \vspace{-3mm}
    \label{fig:gcn_architecture}
    \vspace{-5mm}
\end{figure}
We fuse these two histories together using graph convolution. We compute a degree-normalized adjacency matrix $\hat{\mathbf{A}} = \mathbf{D}^{-\frac{1}{2}}\mathbf{A}\mathbf{D}^{-\frac{1}{2}}$, where $\mathbf{D}$ is the degree matrix of $\mathbf{A}$. This multiplication normalizes the effect of neighboring vertices so that higher degree vertices do not get over-weighted. Now, our convolution at the $m^{th}$ depth is computed as,
\begin{equation}
    \mathbf{H_{m+1}} = \sigma_{r}(\hat{\mathbf{A}} \cdot \mathbf{H_{m}} \cdot \mathbf{K_g^m})
\end{equation}
where $\mathbf{K_g}$ is the graph convolution kernel to be learned while training, and $\mathbf{H_m}$ and $\mathbf{H_{m+1}}$ are the input and the output for the $m^{th}$ convolution respectively. Since we use three consecutive convolution layers, the final feature representation is $\mathbf{H_3}$. 

For predicting whether there will be a conflicting engagement between users $u_i, u_j$, we select the $i^{th}$ and the $j^{th}$ feature vectors of $\mathbf{H_3}$ and compute a score $y\in(0,1)$ as follows:
\begin{equation}
    \mathbf{E} = [\mathbf{H_3}[i],\mathbf{H_3}[i]]
\end{equation}
\begin{equation}
    y = \sigma_s(\mathbf{K_{c}}^\top \cdot \mathbf{E} + \mathbf{B_{c}})
\end{equation}
where $[\cdot,\cdot]$ stands for the concatenation operator, $\mathbf{K_{c}}$ and $\mathbf{B_{c}}$ are the kernel and bias for the classification layer respectively, and $\sigma_s(x)=(1+e^{-x})^{-1}$. The complete architecture of the model is illustrated in Figure \ref{fig:gcn_architecture}.
This model is trained to minimize cross-entropy loss between true and predicted labels.

\subsection{SVM-based frameworks}
Graph convolution automatically learns feature representation for the interaction between user pairs from node features and connectivity of the nodes. For SVM, we need to manually identify interaction features. We extract the following textual and network based features for each user pair $u_i, u_j$:
\begin{enumerate}
    \item {\bf Count of relevant common tokens} from the previous comments of the users; we take the sum of tf-idf values of common unigram and bigrams in the comment history of both the users.
    \item {\bf Conflict vector} $CV_{ij}$ between the pair computed using TD-sentiment vector $TS_D$ following Eq.~\ref{eq:td-senti-vector}; given  previous $N_i^k$ comments of user $u_i$, $\{C_0, C_1,\dots,C_{N_i^k}\}$ where the term $T[k]$ appear, we compute $TS_{u_i}$, the target sentiment vector of $u_i$ averaged over the history as,
    \begin{equation}
        TS_{u_i}[k] = \frac{1}{N_i^k}\sum_{l=0}^{N_i^k}TS_{C_l}[k]
    \end{equation}
    We compute $CV_{ij}$ as the element-wise absolute difference between $TS_{u_i}$ and $TS_{u_j}$.
    \item {\bf Common news sources}, $CN_{ij}$ taken as a vector of length equal to the number of news sources; for news source $k$, $CN_{ij}[k]$ indicates the number of articles from this news source where $u_i, u_j$ both are engaged.
    \item {\bf Common discussions}, indicating the count of discussions where both $u_i$ and $u_j$ are engaged.
    \item {\bf Previous mutual engagement}, the total number of previous interactions between $u_i$ and $u_j$.
    \item {\bf Previous conflict}, the average of mutual conflicts between $u_i$ and $u_j$ for their previous engagements.
    \item {\bf Neighbor interactions}, the count of conflicting and non-conflicting engagements for each user with its neighbor nodes.
\end{enumerate}
We use three SVMs with Gaussian kernel -- first SVM uses  all the features mentioned above (SVM-all), the second one (SVM-text) uses only text based features (features 1 and 2) and the third one (SVM-net) uses only network based features (features 3-5). SVM-net, which has been used for negative link prediction by Wang et al. \cite{wang2015link}, serves as our external baseline.

\section{Experimental Results}
\label{sec:eval}
For the news-user conflict prediction task, total size of our feature vector is $8,136$. On a total set of $41,430$ news articles, we used $80:20$ train-test split keeping the fractions of different news sources same over train and test data.\footnote{We used scikit-learn framework (\url{https://scikit-learn.org/stable/}) to implement all the regression models mentioned.}

For the user-user conflict prediction task, the number of features representing user nodes in the graph convolution model is $8,236$. To construct enclosing subgraphs from user-user engagement network, we set the value of $d_{max}$ (defined in Section ~\ref{subsec:gcn}) to be $100$. This results in adjacency matrices with an upper bound of $5000$ nodes. \footnote{We implement this model using Keras (\url{https://keras.io/}) and Tensorflow frameworks (\url{https://www.tensorflow.org/}).} We perform this prediction on 25 instances of the dynamic user engagement network, taking a total of $1,637$ different subgraphs from these instances. For any user pair on these subgraphs, if there is a conflicting engagement between them over an interval of next 24 hours, we label them as positive, otherwise negative. We take $213,998$ different user pairs altogether, randomly sampling equal numbers of positive and negative labels to avoid bias. Here again, we split the samples into $80:20$ train-test splits, with 15\% of the train data used as the development set to tune the parameters. We use Nadam (Adam with Nesterov momentum) optimization to train the model, with a batch size of $256$.

\subsection{Evaluation of conflict quantification}
We test our conflict measurement on the manually annotated news-comment and comment-comment pairs (Section \ref{sec:data}). To deal with different ranges, we normalize the $cf$ values to the $[0-10]$ interval and measure Root Mean Squared Error (MSE). We also consider ranking comments accordingly to their conflicting tendency towards a particular news article and a particular comment. We compute the Mean Average Precision (MAP) of the ranking and Mean Reciprocal Rank (MRR) for top ranking position based on the ground-truth annotation mentioned in Section \ref{sec:data}.
\begin{table}[!t]
\small
    \centering
    \begin{tabular}{|l|c|c|c|}
    \hline
      {\bf Conflict type} & {\bf RMSE}  & {\bf MAP} & {\bf MRR}\\
    \hline
      News-comment conflict & 0.96 & 0.77 & 0.86 \\
      Comment-comment conflict & 0.79 & 0.83 & 0.91\\
    \hline
    \end{tabular}
    \caption{Evaluation of conflict measurement on manually annotated conflict ratings.}
    \label{tab:conflict_evaluation}
    \vspace{-5mm}
\end{table}
\par As observed in Table \ref{tab:conflict_evaluation}, measuring inter-comment conflict is rather an easier task compared to news-comment conflict. The feedback obtained from the annotators reveal that as most news articles are written in an objective style with less explicit opinion, it is hard to apprehend whether a comment holds opposite opinion to the news.
\begin{figure}
    \centering
    \includegraphics[width=0.8\columnwidth]{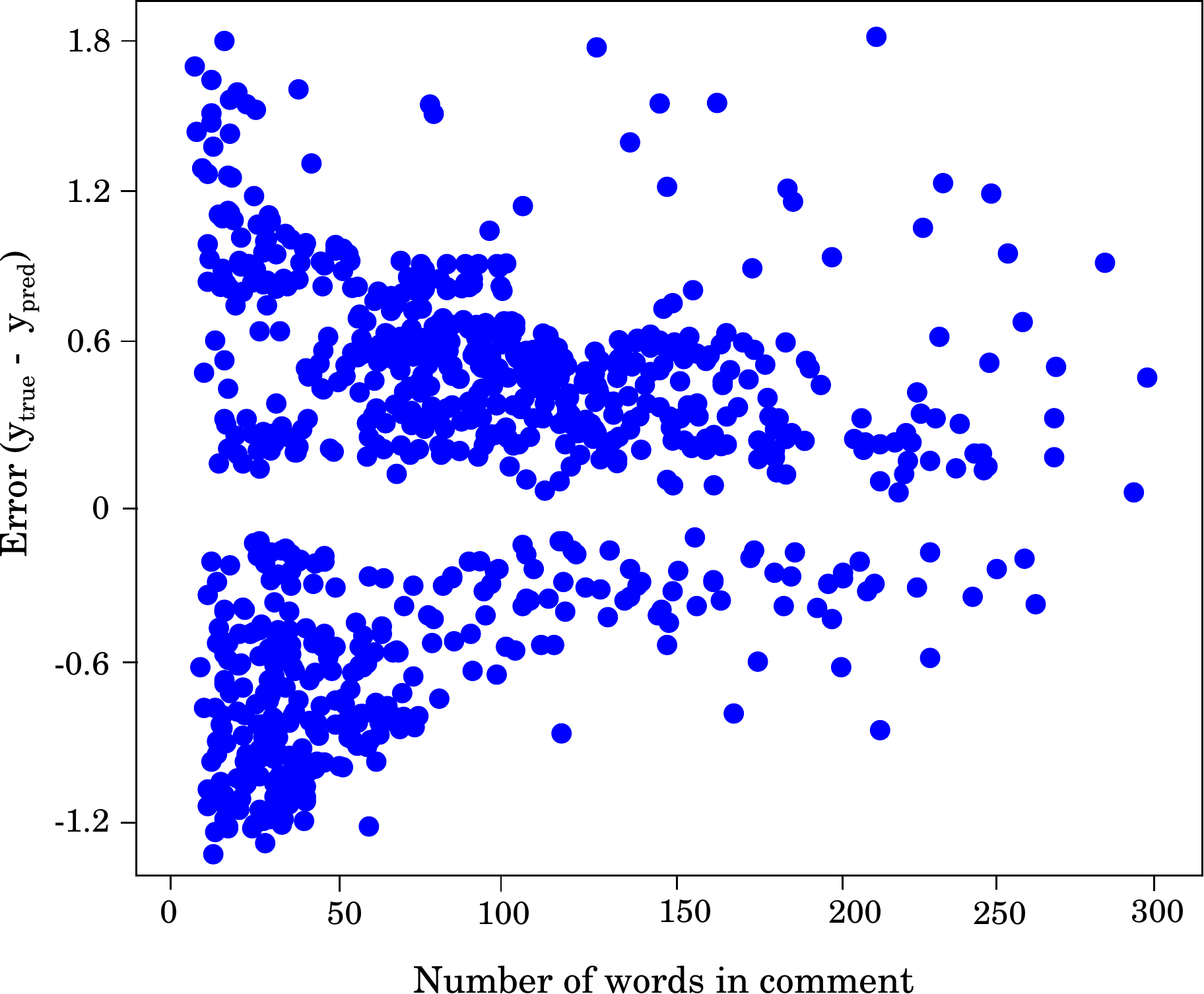}
    \caption{Error in conflict score vs. size of comments in words.}
    \label{fig:error_vs_size}
    \vspace{-5mm}
\end{figure}

As there is no previous work in quantifying conflict between two text documents over online discussions, we implement the agreement-disagreement detection models proposed by Rosenthal and McKewon \cite{rosenthal2015couldn} ({\bf Baseline-I}) and Dutta et al. \cite{dutta2019did} ({\bf Baseline-II}). Baseline-I performs a three-class classification: {\em agreement}, {\em disagreement} and {\em none}. We identify disagreement as conflict and rest of the classes as non-conflict. We also define the probability of the disagreement class (predicted by Baseline-I) for an interaction as a unit norm score of conflict. Similarly, Baseline-II performs a ten-class classification of discourse acts, from which we identify the classes {\em disagreement} and {\em negative reaction} together as conflict, and rest of the classes as non-conflict. Sum of the probabilities of these two mentioned classes is defined as unit norm conflict score predicted by Baseline-II.

We compare our strategy of conflict score prediction
with the baselines through a three-way evaluation strategy: 
\begin{enumerate}
    \item We define a binary classification of interactions into conflict and non-conflict, evaluated using ROC-AUC;
    \item We define a regression of the degree of conflict, where we scale the outputs of each model to the interval $[0,10]$ and evaluate using RMSE;
    \item We define a ranking problem of the interactions according to their degree of conflict, and evaluate using MAP.
\end{enumerate}

As both the baselines perform their corresponding tasks (stance classification and discourse act classification) on discussion data, we perform this comparisons only for the comment-comment conflict prediction.
\begin{table}[]
\small
    \centering
    \begin{tabular}{|c|c|c|c|}
    \hline
        {\bf Metric} & \specialcellC{\bf Our method\\ {\bf (conflict factor)}} & {\bf Baseline 1} & {\bf Baseline 2}\\
    \hline
        AUC & 0.79 & 0.79 & 0.62 \\
        MAP & 0.83 & 0.61 & 0.55 \\
        RMSE & 0.79 & 1.67 & 2.09 \\\hline
    \end{tabular}
    \caption{Comparison of conflict score with baselines.}
    \label{tab:conflict_baseline}
    \vspace{-8mm}
\end{table}

 Table \ref{tab:conflict_baseline} shows that our proposed strategy outperforms both the baselines for ranking and regression tasks. This is quite expected as both the baseline models are actually classification frameworks. For the binary classification of conflicting and non-conflicting interactions, our strategy ties with Baseline-I.

Figure \ref{fig:error_vs_size} plots the variance of error in conflict score with the change in the comment length. For news-comment pairs, we only take the comment length, while for comment-comment pairs we take the average of the length of both comments. To see whether the error in our score has any bias towards underestimation / overestimation, we take the difference $(y_{true}-y_{pred})$, where $y_{true}$ and $y_{pred}$ are manually annotated score and computed $cf$ respectively. As we can see in Figure~\ref{fig:error_vs_size}, our computed score underestimates conflict when comments are short, and overestimates as the size grows (more negative errors for size approximately less than $60$ words; more positive errors afterwards). Also, absolute error rate decreases with increasing size of comments. 

Such error pattern can be explained from the definition of conflict measurement itself. We use the sum of the absolute differences of sentiment towards specific targets common in documents, as conflict score which increases with the number of common targets present. As the length of the comments increases, the common word set also increases, and small differences add up to large conflict scores. For short comments, the number of common targets are also small, and the score tends to reflect less conflict than actual. For shorter comments, another problem is the use of semantically similar words occurring as targets in any of the comments in a given  pair. For example, the sentences  `\textit{We do not support Democrats}' and `\textit{We support Hilary}' are actually conflicting, as the targets \textit{Hilary} and \textit{Democrats} are semantically similar. But due to no common words, these pairs will be identified as non-conflicting.

However as our dataset suggests, the fraction of comments having greater than $50$ words is $0.79$; and the ratio between the number of words and targets is $17.678$. This is particular to the online discussion forums, where users tend to get engaged in an elaborate manner, and therefore reduces the error margin of our conflict score. Our model achieves $0.96$ and $0.79$ RMSE for news-comment and comment-comment conflict, respectively, over the interval $[0,10]$ which might be considered as significantly accurate for conflict modeling.

\subsection{Evaluation of news-user conflict prediction}
\begin{table}[]
    \centering
    \begin{tabular}{|l|c|c|c|}
    \hline
        {\bf Model} & {\bf MSE} & {\bf RMSE} & {\bf sMAPE} \\
        \hline
        Random Forest & 6.194 & 2.489 & 0.099 \\
        SVR & 4.041 & 2.010 & {\bf 0.077} \\
        Lasso & {\bf 3.179} & {\bf 1.783} & 0.080 \\
        \hline
    \end{tabular}
    \caption{Performance of different regression algorithms for news-user conflict prediction.}
    \label{tab:performance_news}
    \vspace{-8mm}
\end{table}
In our dataset, the news conflict scores (computed using Eq.~\ref{eq:news_conflict_score}) of the news articles vary from 0 to $138.15$. In Table \ref{tab:performance_news}, we present the MSE (Mean Squared Error), RMSE and sMAPE (Symmetric Mean Absolute Percentage Error) for predicting news conflict scores using different regression algorithms. In terms of MSE and RMSE, Lasso regression performs the best, while SVR is the best performing one when evaluated using sMAPE.  
\begin{figure}
    \centering
    \includegraphics[width=0.4\textwidth]{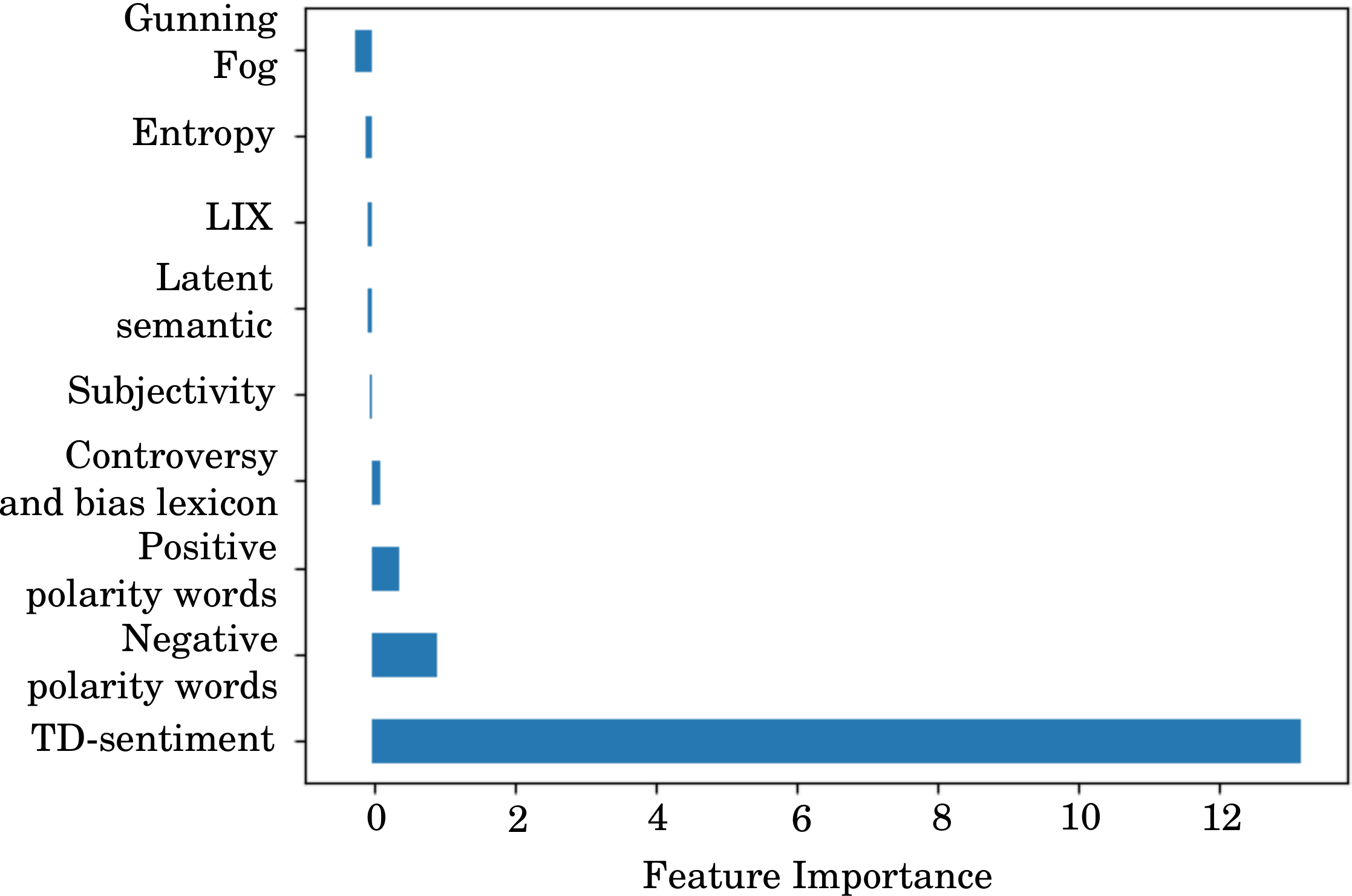}
    \vspace{-3mm}
    \caption{Importance of different features for news-user conflict prediction.}
    \label{fig:feature_importance}
    \vspace{-5mm}
\end{figure}
\par We check to see which features are given more importance by our best performing regression algorithms. As we can see in Figure \ref{fig:feature_importance}, term dependent sentiments are the most useful ones to predict how much likely is a news article to get negative feedback. In fact, this feature achieves way more importance compared to its next competitor, which again are polarity-oriented features. Interestingly,  the count of negative polarity words has higher importance than the count of positive polarity words. The high importance of polarity related features may signify that news report expressing polarized bias tends to get more conflicting remarks. Readability indices (Gunning-Fog and LIX), albeit low, play some role in the prediction task. In fact, Gunning-Fog is substantially more useful compared to LIX.

\subsection{Evaluation of inter-user conflict prediction}

We evaluate all four  models for two cases:  (i) whole of the test data where a pair of users may or may not have previous interaction history, and (ii) user pairs who have no interaction history before the prediction instance. We present the evaluation results in Table \ref{tab:user-user_evaluation}. For the whole test data, SVM model with all the features performs the best. It is readily conclusive that, network-based features are of greater importance compared to text-based features for this task. 

However, when there is no previous interaction history between two users, graph convolution beats all the models by a substantial margin. In fact, when there is no previous engagement history between  users, the only feature available to the SVM model is the neighbour interactions; which means SVM-all and SVM-net actually become the same model, and SVM-text becomes a model with all zero features with all zero output. 
\section{Conflict dynamics}
\label{sec:conflict_dynamics}

We introspect into the dynamics of conflict in \textit{r/news} community using the conflict measurements that we propose in Eq.~\ref{eq:conflict_factor} (for inter-user conflict) and Eq.~\ref{eq:news_conflict_score} (for an aggregate conflict that a news article receives from the users).  

\begin{table}[!t]
    \centering
    \begin{tabular}{|l|c|c|c|c|}
    \hline
        {\bf Evaluation} & {\bf SVM-all} & {\bf SVM-text} & {\bf SVM-net} & {\bf GCN}\\
    \hline
        Acc. & {\bf 0.89} & 0.64 & 0.85 & 0.87\\
        AUC & {\bf 0.89} & 0.62 & 0.84 & 0.86\\
        Acc. (new) & 0.67 & 0.43 & 0.67 & {\bf 0.72}\\
        AUC (new) & 0.65 & 0.43 & 0.65 & {\bf 0.69}\\
    \hline
    \end{tabular}
    \caption{Evaluation of all the models for user-user conflict prediction. Accuracy is abbreviated as Acc. Acc. (new) and AUC (new) signify evaluation results for user pairs with no previous interactions.}
    \label{tab:user-user_evaluation}
    \vspace{-5mm}
\end{table}

\subsection{Patterns of conflict for different news sources}
\begin{figure}
    \centering
    \includegraphics[width=0.35\textwidth]{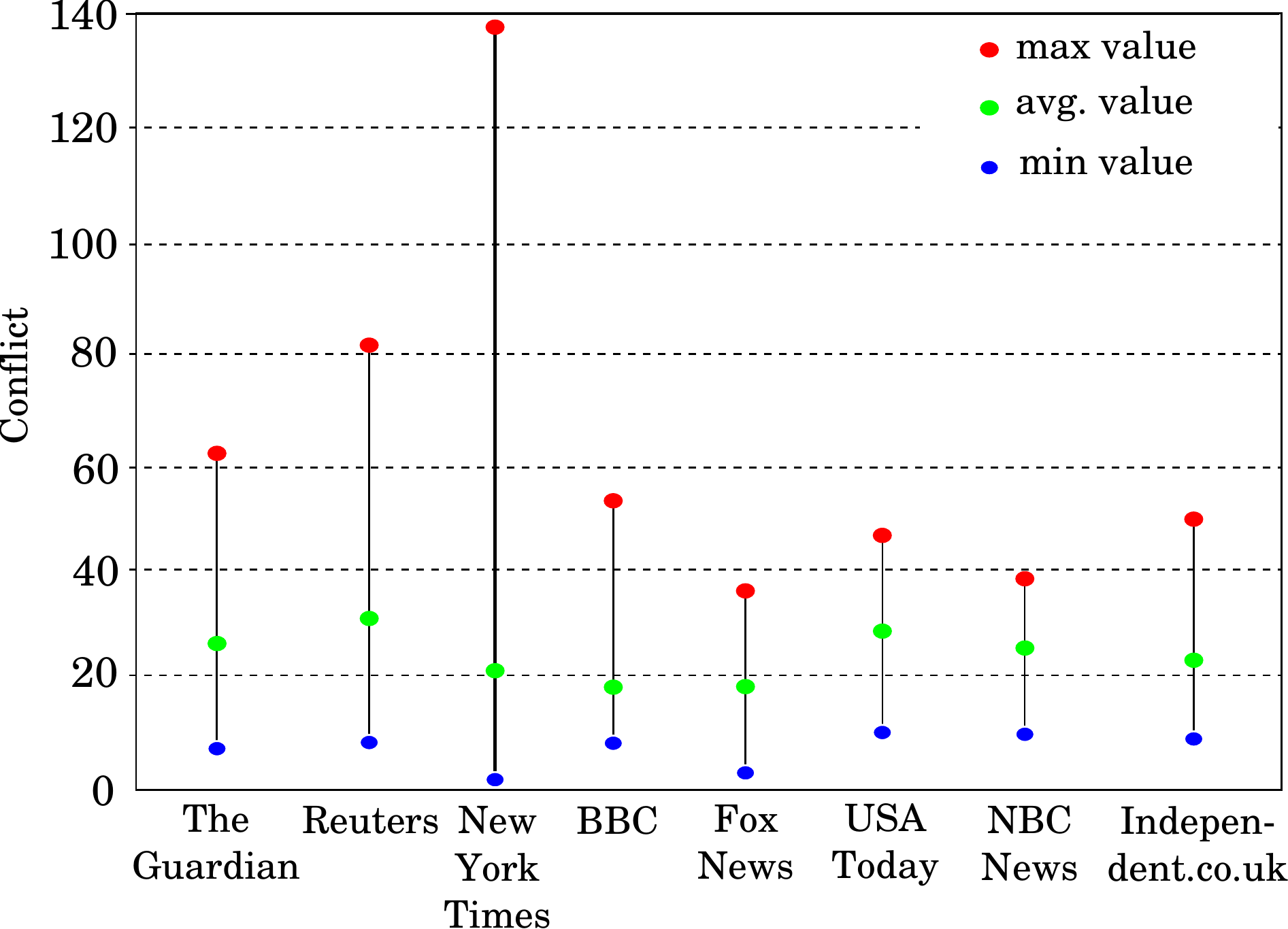}
    \vspace{-2mm}
    \caption{Distribution of maximum, minimum and average conflict scores for different news sources. This plot is for only top 7 news sources (ranked by number of articles).}
    \label{fig:max_min_news_conflict}
    \vspace{-5mm}
\end{figure}

Different news sources tend to face different degree of conflict from the users. In Figure \ref{fig:max_min_news_conflict}, we plot maximum, minimum and average news conflict for different news sources in our dataset. Although the average conflict for different sources is in a comparable range, maximum values vary greatly. News sources such as \textit{Fox News}, \textit{USA Today} or \textit{NBC News} maintain a sustained negative response, whereas \textit{New York Times} or \textit{Reuters} provoke sharp outrage at the some point. 

\begin{figure}
    \centering
    \includegraphics[width=0.46\textwidth]{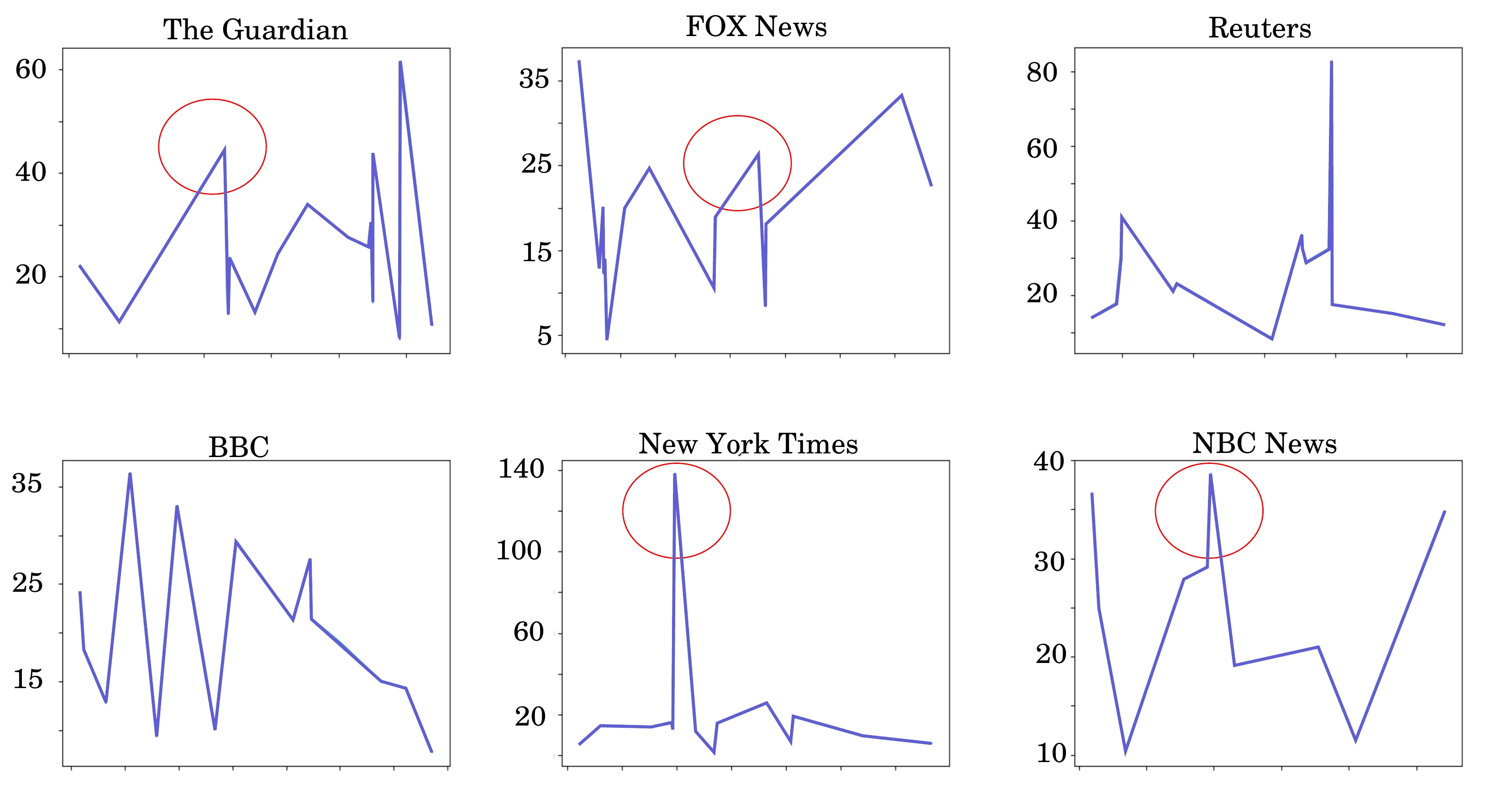}
    \vspace{-5mm}
    \caption{Temporal variation of news-user conflict for various news sources; conflict score and time are represented in y- and x-axis respectively. All the plots have time frame starting from Nov 17 - Dec 28, 2017. Red circled peaks denote rise in conflict due to articles corresponding to a particular event.}
    \label{fig:temporal_conflict}
    \vspace{-5mm}
\end{figure}

\par We find that this outrage is signified by an article published in NYTimes on Dec 1, 2017, titled \textit{Michael Flynn Pleads Guilty to Lying to the F.B.I. and Will Cooperate With Russia Inquiry}\footnote{\url{https://www.nytimes.com/2017/12/01/us/politics/michael-flynn-guilty-russia-investigation.html}}. Figure \ref{fig:temporal_conflict} also indicates the sharp peak for New York Times corresponding to this article. The Guardian, Fox news and NBC News have similar peaks (red-circled) at nearby time instance, all corresponding to articles related to the same event. One can draw an intuitive correlation between the  posting time of the article in the forum and the rise in conflict. It is important to note that at the time of posting, we identify the time when the news appeared on Reddit, not the time of its appearance on web. 

\begin{figure*}[t]
    \centering
    \includegraphics[width=0.9\textwidth, trim=1.4cm 0cm 1.6cm 0cm,clip]{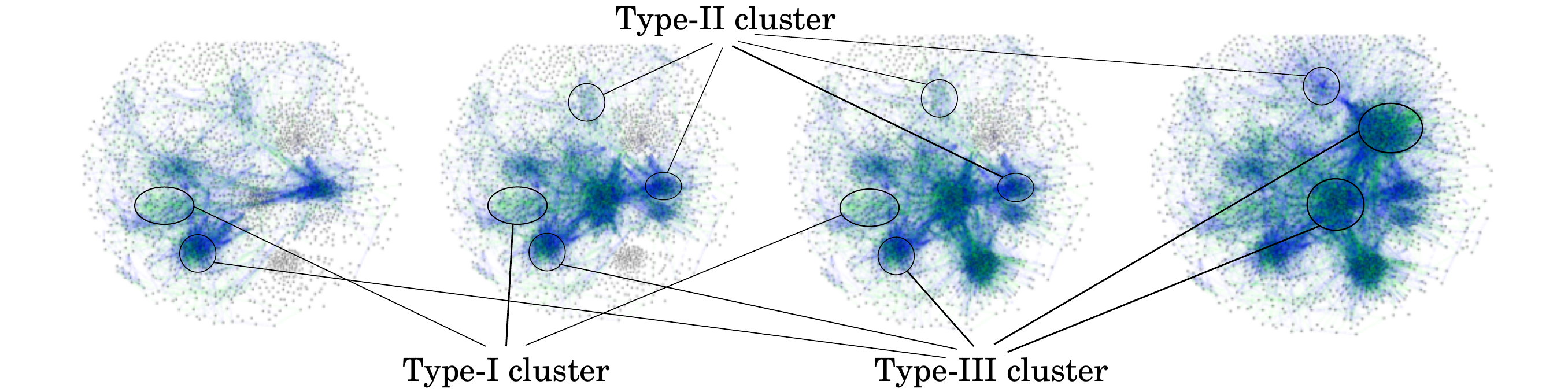}
    \vspace{-3mm}
    \caption{Snapshots of cluster formation in user-user engagement graph (left to right); blue and green edges correspond to controversial and non-controversial engagements respectively.}
    \label{fig:conflict_community}
    \vspace{-5mm}
\end{figure*}

\subsection{Engagement dynamics and inter-user conflict}
To explore how conflict effects user engagement over \textit{r/news}, we construct a temporal graph $G'(t)=\{V'(t),E'(t)\}$, where $v_i(t_i)\in V'(t)$ corresponds to user $u_i$ who is engaged in a discussion at time $t_i$ for the first time. For every pair of users $(u_i,u_j)$ engaging with each other (anyone of them commenting in reply to the other) at time $t_{ij}$, there is an edge $e_{ij}(t_{ij})\in E'(t)$. For better visualization, we classify edges as conflicting (blue) and non-conflicting (green), and plot only a subgraph using $5000$ vertices. We use Fruchterman Reingold layout algorithm \cite{fruchterman1991graph} on Gephi \cite{bastian2009gephi} to plot the graph and DyCoNet\cite{kauffman2014dyconet} to identify communities. In Figure \ref{fig:conflict_community}, we present snapshots of the evolving graph. Each snapshot is taken at a time difference of approximately 24 hours, presenting a 4-day long abstraction through this engagement subgraph. 

We can observe the formation of separate user clusters in terms of engagement. It is interesting to see that there are some clusters where users are predominantly engaged with each other in a conflicting manner (blue regions) and some in a non-conflicting manner (green regions). We also identify three different types of engagement patterns in user clusters:
\begin{itemize}
    \item {\bf Type-I} clusters tend to be formed with non-conflicting engagement between users. Users in these clusters do not seem to get engaged in conflicting manner with users in other clusters as well. 
    \item {\bf Type-II} clusters are formed with users having mutual conflict. They tend to have conflicting interactions with other clusters as well.
    \item {\bf Type-III} clusters show a organization-like behavior. These users maintain almost non-conflicting engagement with each other, but aggressive towards other clusters (mostly green regions inside the cluster and blue ones outwards in Figure \ref{fig:conflict_community}).
\end{itemize}

Type-III clusters tend to grow most compared to type-I and type-II clusters. Different type-III clusters have most inter-cluster conflicts, even greater than that of type-II clusters. Type-I clusters show least growth rate among   three types, signifying that these users are less prone to go out of their `comfort zone'. 

This cluster types are of course not completely rigid. Although there is no sign of conversion between type-I and II, both of them can slowly convert into type-III. It is intriguing to observe two different patterns in the formation of type-III cluster -- (i) Some of them emerge as type-III from the beginning. Users having no previous engagement form non-conflicting connections with each other. This may signify a probable community interaction among them beyond the discussion platform such as organized campaigners, small group of people using multiple fake user accounts {\em aka} sockpuppets \cite{Kumar:2017}, people are accustomed to each other in real life and sharing similar opinions, etc. (ii) Some of them started as type-I or II and slowly get converted into type-III, which  possibly signifies the evolution of engagement via predominant platform interaction. Users in type-II clusters start changing opinion towards each other with long term interaction and get converted into type-III. Similarly, type-I users tend to start interacting with opposite opinions and convert themselves into type-III. We observe that 33\% of the type-III clusters at the end of time frame are the ones converted from type-II, whereas 48\% are from type-I. Rest of them started growing as type-III clusters.
\begin{figure}[h]
    \centering
    \includegraphics[scale=0.4]{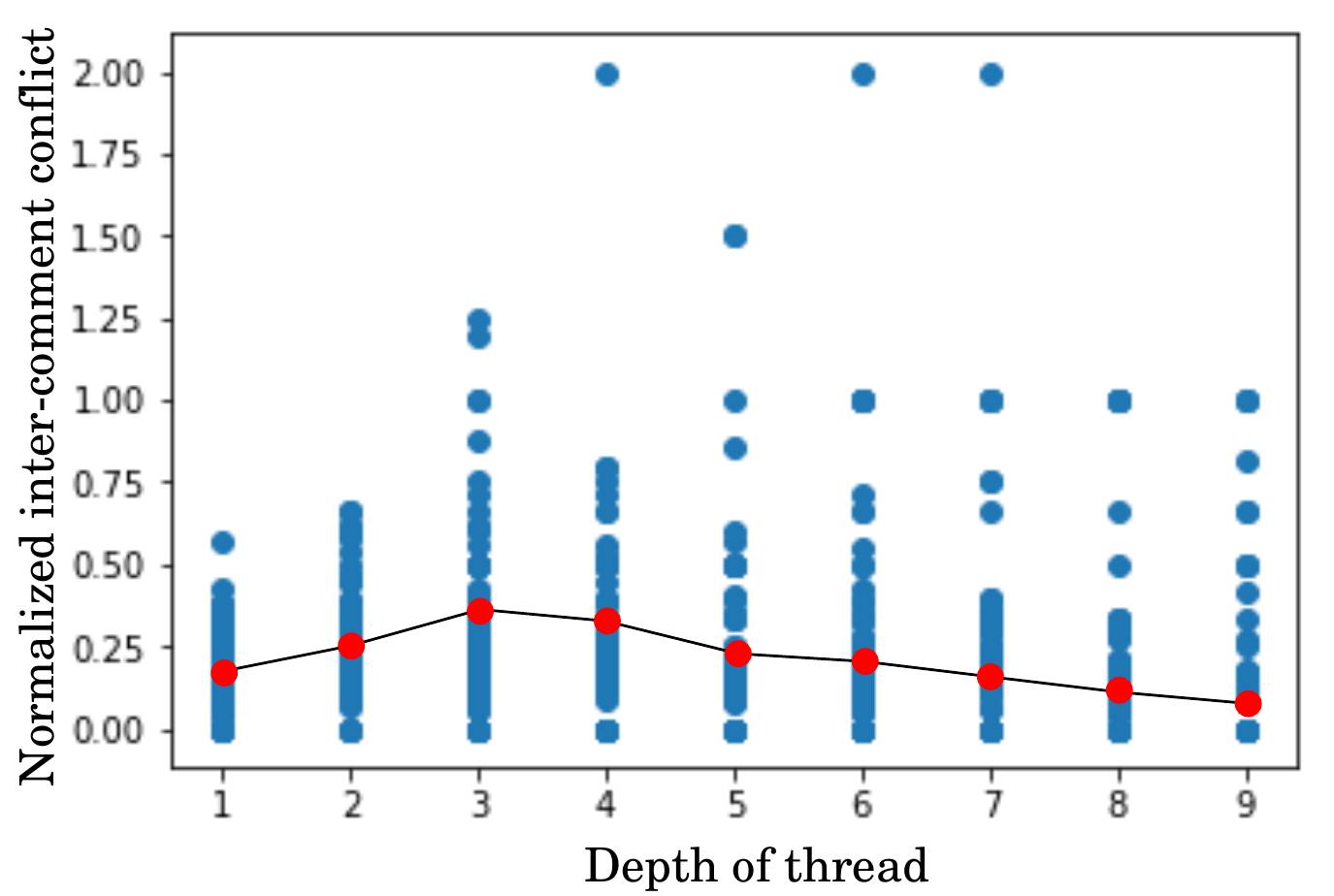}
    \vspace{-3mm}
    \caption{Variation of inter-comment conflict with depth of comments in discussion tree.}
    \label{fig:depth_conflict}
    \vspace{-4mm}
\end{figure}

Formation and evolution of these clusters closely follow the abstract model of user engagement in Figure \ref{fig:conflict_state}. A repeated transition from state 1 along the self loop results in type-I cluster, whereas the same  happening on state 2 will result in a type-II cluster. If all the user pairs from state 1 start conflicting with each other, it will lead to a transition to state 2, which implies that a type-I cluster is transformed into type-II. This can only be  possible  hypothetically; however we did not find any such evidence in our dataset. Likewise, transition from state 1 or 2 to state 3 signifies preferential conflict, resembling type-III clusters.
\begin{table*}[]
\scalebox{0.8}{
\begin{tabular}{|c|l|l|l|}
\hline
                 \specialcellC{{\bf Cluster index}\\{\bf ranked along size}} & \multicolumn{1}{c|}{{\bf Instance 1}} & \multicolumn{1}{c|}{{\bf Instance 2}} & \multicolumn{1}{c|}{{\bf Instance 3}} \\ \hline
\multirow{4}{*}{1} & Baltimore News (40.02\%) & Comic Book (76.48\%) & New York Times (49.03\%) \\
                  & Wichita Eagle (25.92\%) & Wichita Eagle (11.89\%) & Fox News (25.08\%) \\
                  & National Geographic (20.00\%) & Fox News (1.91\%) & abc13 (25.88\%) \\
                  & Fox News (13.33\%) & Detroit News (0.54\%) &  \\ \hline
\multirow{3}{*}{2} & Baltimore News (100\%) & Wichita Eagle (50.57\%) & BBC (78.52\%) \\
                  &  & Baltimore News (47.12\%) & Independent (18.61\%) \\
                  &  & National Geographic (2.29\%) &  New York Times (2.87\%)\\
                  \hline
\multirow{3}{*}{3} & abc13 (44.44\%) & Baltimore News (100\%) & Guardian (42.98\%)\\
                  & Fox News (27.78\%) &  & Independent (41.32\%) \\
                  & Baltimore News (22.24\%) &  & Detroit News (15.70\%) \\
                  \hline
\end{tabular}}
\caption{Percentage of different news sources in user clusters of user-user engagement network. We show the statistics of three largest clusters at three different instances of the network. Up to top four news sources (according to \%-contribution) is shown.}
\label{tab:source_cluster}
\vspace{-8mm}
\end{table*}

In Figure \ref{fig:depth_conflict}, we plot the variation of inter-comment conflict with the depth of the comments in discussion thread tree. We normalize conflict scores to $(0,2)$ interval. For comment pairs at depth $i$ and $i+1$, we plot their conflict at depth $i$. As it is evident from the plot, a discussion thread is most prone to conflict at depth levels 3 and 4. For interactions at more depth, variance goes up substantially, but average inter-conflict score drops steadily.

Table \ref{tab:source_cluster} shows an example statistics of different news sources regarding which discussions lead to user clusters. We report this for three different instances $G'(t_1)$, $G'(t_2)$, and $G'(t_3)$ at time $t_1$, $t_2$, and $t_3$ respectively. We take the discussions initiated within past 24 hours for each instance of the network and map the users in each of the largest three clusters to those discussions. As each discussion is related to a news source, this finally maps news sources to clusters. As we can see in Table \ref{tab:source_cluster}, there are several common news sources present in first and second instances, whereas almost no common sources is found in the third instance. 

\vspace{-3mm}
\section{Conclusion}

In this paper, we studied conflict dynamics over online discussions inside Reddit {\em r/news} community. We proposed a novel, continuous-valued quantification of inter-document conflict. Using this measurement we attempted to predict how much negative response a news article is going to face from audience in online discussion platforms, solely based on its textual features. We proposed an SVM based model and a graph convolutional model to predict future conflict between pairs of users. Extensive evaluation showed that network-based features are more important in conflict link prediction compared to textual content-based features.

Our analyses provide novel insights into the conflict dynamics over large-scale online discussion. We show how different news sources get different reactions from their audience and how this varies temporally. We identified three distinct types of user clusters developed in Reddit {\em r/news} community, based on the attitude towards other users and engagement patterns. We also provided a hypothetical state-transition model of user engagement, which is closely followed by actual interaction patterns.

\section*{Acknowledgement}
The project was partially supported by Ramanujan Fellowship (SERB, India),  Early  Career  Research  Award  (ECR/2017/001691), the Infosys Centre for AI, IIITD, and State Government Fellowship, Jadavpur University.

\bibliographystyle{ACM-Reference-Format}
\bibliography{bibliography}
\end{document}